\documentclass{elsarticle}
\usepackage{natbib}
\usepackage{amsmath}
\input{tcilatex}

\begin{document}

\begin{frontmatter}
\title{Estimating a Signal In the Presence of an Unknown Background}
\author[wolf]{Wolfgang A. Rolke}
\address[wolf]{Department of Mathematics, University of Puerto Rico - Mayag\"{u}ez, Mayag\"{u}ez, PR 00681, USA, 
\newline Postal Address: PO Box 3486, Mayag\"{u}ez, PR 00681, 
\newline Tel: (787) 255-1793, Email: wolfgang@puerto-rico.net}
\author[ang]{Angel M. L\'{o}pez}
\address[ang]{Department of Physics, University of Puerto Rico - Mayag\"{u}ez, Mayag\"{u}ez, PR 00681, USA}
\begin{abstract}
  We describe  a method for fitting distributions to data which only requires knowledge of the parametric form of
either the signal or the background but not both. The unknown distribution is fit using a non-parametric kernel density estimator. 
A transformation is used to avoid a problem at the data boundaries.
The method returns parameter estimates as well as errors on those estimates. Simulation studies show that these estimates are unbiased
and that the errors are correct. 
\end{abstract}
\end{frontmatter}\newpage

\section{Introduction}

Almost all data sets encountered in High Energy Physics (HEP) have events
that were generated by different mechanisms. Interest often focuses on one
of these mechanisms, what often is called a signal. By judicious cutting on
auxiliary variables it is often possible to bring out the signal, that is,
to improve the signal to noise ratio, but it is usually not possible to
eliminate the noise, or background events, altogether. In many cases the
researchers want to estimate the parameters involved in the signal such as
the location and the width of an invariant mass peak, together with their
statistical errors. In order to do so, they need to find a model for the
data, and because the data still contains background events the model
usually has the form of a mixture distribution. In practice it often turns
out to be fairly straightforward to find a parametric description for the
signal density, for example from theoretical considerations. In contrast it
is often very difficult to model the background density, and the typical
approach is one of trial and error: fit a number of shapes until a
satisfactory one is found. This approach has several drawbacks: (i) it is
quite time consuming, (ii) different researchers might end up with different
parametrizations, (iii) it is hard to know when to stop in order not to
overfit the data and (iv) it is almost impossible to gauge the systematic
uncertainty of an incorrectly specified background on the parameter
estimates.

There are, however, other solutions to the problem of estimating a density,
namely the so-called nonparametric density (NPD) estimators. A number of
such methods are known, such as kernel methods, nearest neighbor methods,
the method of penalized likelihood and others. We could apply any of these
methods to HEP data but unfortunately they do not yield the parameters that
are the ultimate goal of the fitting process.

The semiparametric method that will be discussed here combines traditional
parametric fitting with nonparametric density estimation. A parametric
description is used for the signal, the background is modeled
nonparametrically and these two ingredients are combined into one fitting
function.

This method is made especially useful by two features of HEP data. First, we
often have available a sample of pure background events, either by
eliminating potential signal events by cutting on auxiliary variables or
from Monte Carlo. This will allow us to employ the most powerful techniques
developed for NPD estimation, for example the methods for choosing the
optimal amount of smoothing. Secondly, data in HEP is usually supported on a
finite interval. This normally would lead to the so-called boundary problem
in NPD density estimation, a seemingly innocuous problem for which to date
no single solution is known. In this work we solve this problem by
transforming the data, a solution which is viable because in HEP we are not
interested in the NPD estimate itself but in the signal parameters. As our
simulation studies show this solution of the boundary problem can be applied
in a wide variety of situations and leads to correct estimates of the
parameters and their errors.

There is a large volume of literature in the field of Statistics on
nonparametric density estimation. Standard references include Scott \cite%
{Scott}, Silverman \cite{Silverman}, Tapia and Thompson \cite{Tapia}, Fryer 
\cite{Fryer} and Bean \cite{Bean}. Kernel density estimation was first
introduced independently by Rosenblatt \cite{Rosenblatt} and Parzen \cite%
{Parzen}.

One of the early discussions of NPD estimation in HEP was by Cranmer \cite%
{Cranmer1} in 2001, who also discussed their use in multivariate analysis 
\cite{Cranmer2}. It has since been used in a variety of HEP analyses (\cite%
{Lep}, \cite{Babar}, \cite{CDF}, \cite{SNO}).

\section{Kernel Density Estimation}

Nonparametric density estimation is concerned with estimating a probability
density without assuming a functional form for the density. The
\textquotedblleft nonparametric\textquotedblright\ in the name is slightly
misleading because these estimators have in fact something like parameters.
There are a number of different methods that have been studied in some
detail, such as nearest neighbor estimators, splines, orthogonal series, and
so on. Even a histogram (properly scaled) can be viewed as a NPD estimator.
We will concentrate on the best known and most widely used method, the
kernel density (KD) estimator.

Let $X_{1},..,X_{n}$ be observations from some unknown density $f$. Let $K$
be any continuous, nonnegative and symmetric function with $K(x)\rightarrow
0 $ as $x\rightarrow \pm \infty $. Often $K$ is chosen to be a probability
density itself, say a Gaussian density. We then get the following definition
of a nonparametric kernel density estimator: 
\begin{equation}
\widehat{f}_{h}(x)=\frac{1}{nh}\sum_{i=1}^{n}K(\frac{x-X_{i}}{h})
\label{kernel dev}
\end{equation}

As has been noted many times, the choice of the kernel function is of minor
importance (Scott \cite{Scott}). We will use the Gaussian density in this
work but other kernels could also be chosen.

The $h$ is a tuning parameter called the bandwidth. It governs the
variance-bias trade-off, with a small value of $h$ resulting in a very
ragged curve whereas a large value oversmoothes the density.

How do we choose the bandwidth $h$? This has been one of the most active
areas of research in the field of Statistics in the last 20 years, and a
number of methods have been proposed. To begin with a criterion for
\textquotedblleft best\textquotedblright\ is needed, and most research has
focused on either the \emph{Integrated Squared Error} 
\begin{equation}
ISE(h)=\int_{-\infty }^{\infty }(\widehat{f}_{h}(x)-f(x))^{2}dx
\end{equation}%
or the \emph{Mean Integrated Squared Error} $MISE(h)=E[ISE(h)]$, the
expected value of $ISE(h)$. Clearly both depend on the unknown density $f$
and therefore have to be estimated from the data. Another criterion is the 
\emph{Asymptotic Mean Integrated Squared Error} $AMISE(h)$, based on a
consideration of the large sample behavior of the $MISE$. Alternatively one
can consider the corresponding \emph{Mean Integrated Absolute Value Error}
criterion (Devroye and Gy\"{o}rfi \cite{Devroye}). Unfortunately, there is
no obvious way to choose among these criteria, although Jones \cite{Jones1}
gives some strong arguments in favor of $MISE(h)$. Detailed discussions on
bandwidth selection can be found in Turlach \cite{Turlach} and Chiu \cite%
{Chiu}. In this paper we will not advocate the use of one specific criterion
but rather we will vary $h$ over a range of reasonable values, thereby
gaining insight into the systematic uncertainty introduced by this choice.

It is intuitively obvious, and can be shown to be mathematically correct,
that, in regions where the true density changes slowly, the optimal value of 
$h $ should be larger than in regions where $f$ changes rapidly. It is
therefore reasonable to use an adaptive bandwidth as follows:%
\begin{equation}
\widehat{f}_{h_{x}}(x)=\frac{1}{nh}\sum_{i=1}^{n}K(\frac{x-X_{i}}{h_{x}})
\end{equation}%
One suggestion is to use a fixed bandwidth $h$ to get a pilot estimate $%
\widehat{f}_{h}$ and then recompute the density estimate using $h_{x}=h/%
\sqrt{\widehat{f}_{h}(x)}$ (Abramson \cite{Abramson}). In our simulation
studies there was very little difference between the adaptive method and
using a fixed bandwidth, but it is a good idea to investigate this issue in
any specific situation.

For the simulation studies shown later we used a simple bandwidth formula
suggested by Scott \cite{Scott}: let $s$ be the standard deviation of the
data and let $IQR$ be the interquartile range, then 
\begin{equation}
h=1.06\min \{s,IQR/1.34\}n^{-1/5}  \label{bandwidth}
\end{equation}

The extension to multivariate data is straightforward and usually done using
a product kernel:%
\begin{equation}
\widehat{f}_{h_{1},..,h_{d}}(x_{1},..x_{d})=\frac{1}{nh_{1}..h_{d}}%
\sum_{i=1}^{n}\dprod\limits_{j=1}^{d}K_{j}(\frac{x_{j}-X_{j}^{i}}{h_{j}})
\end{equation}

\section{The Problem of Boundaries}

One typical feature of the data sets in HEP is that they are truncated,
i.e., the data are bounded. For univariate data the boundary consists simply
of the two endpoints of an interval $[A,B]$. Unfortunately nonparametric
density estimators have a difficult time dealing with boundaries. The reason
for this is that the lack of data beyond the boundary leads to a smaller
estimate just inside the boundary and, because they have to integrate to $1$%
, they overestimate the density in the middle.

A number of methods have been proposed to deal with the boundary problem.
One approach is to modify the kernel near the boundary (Gasser, M\"{u}ller
and Mammitzsch \cite{Gasser} , Jones \cite{Jones2}). The most serious
problem with these modified boundary kernel methods is that they often lead
to negative density estimates.

Another idea is to fit local polynomials of low order as discussed in Cheng,
Fan and Marron \cite{Cheng}. The argument is that local polynomial
estimation would automatically correct for boundary effects in regression,
and should therefore also work in density estimation (Fan and Gijbels \cite%
{Fan}). A boundary correction, though, takes place only if the polynomial is
of the \textquotedblleft correct\textquotedblright\ order, else it can make
the boundary effect even larger. Local polynomial fitting has not been as
successful in density estimation as it is in regression problems, although
Zhang and Karunamuni \cite{Zhang1} improved the performance of this method
by combining it with a bandwidth variation function.

Another set of boundary correction methods modifies the bandwidth near the
boundaries. The basic idea is that a larger bandwidth close to the boundary
should compensate for the lack of data beyond it. Specific proposals can be
found in Rice \cite{Rice}, Gasser \cite{Gasser} and M\"{u}ller \cite{Muller}%
. In contrast Dai and Sperlich \cite{Dai} actually advocate a smaller
bandwidth.

One of the early attempts at solving the boundary problem was the reflection
method, introduced by Schuster \cite{Shuster} and Silverman\ \cite{Silverman}
and later extended by Cline and Hart \cite{Cline}. A more recent extension
of this method is to create pseudo data to correct for edges (Cowling and
Hall \cite{Cowling}.) This method is more adaptive than the common data
reflection approach in the sense that it corrects also for discontinuities
in derivatives of the density. There is also a whole class of transformation
methods (Wand, Marron and Ruppert \cite{Wand}, Ruppert and Marron \cite%
{Ruppert}, and Yang \cite{Yang}). In addition, Zhang, Karunamuni and Jones 
\cite{Zhang2} suggested a method of generating pseudo data combining the
transformation and reflection methods.

Unfortunately, despite the numerous methods that have been proposed, none
has been shown to yield acceptable results in a wide range of cases. In this
paper we will take advantage of the fact that our ultimate goal is not
density estimation but the estimation of parameters such as the number of
signal events. We can therefore transform the data to the whole real line
and simply do the fitting there. Note that these are transformations of the
data that leave the parameters unchanged.

Say the original observations are $X_{1},..,X_{n}$ and they are located in
the interval $[A,B]$. Then we will calculate new observations $%
X_{1}^{T},..,X_{n}^{T}$ with $X_{k}^{T}=H(X_{k})$; $k=1,..,n$. The
transformation $H$ can be any function that is monotonically increasing,
continuous and that maps $[A,B]\rightarrow (-\infty $,$\infty )$. We will
use $H(x)=\log \left( \frac{x-A}{B-x}\right) $.

When we transform the data, what happens to the density of the original data
and to its parameters? Say the original density was given by $f(x;\theta )$
where $\theta $ is the vector of parameters. Let's denote the density of the
transformed data by $g$. Then 
\begin{equation}
g(y;\theta )=f(H^{-1}(y);\theta )\cdot \frac{d}{dy}H^{-1}(y)  \label{dens1}
\end{equation}%
For the transformation above we have 
\begin{equation}
H^{-1}(y)=\frac{A+Be^{y}}{1+e^{y}}\text{ and }\frac{d}{dy}H^{-1}(y)=\frac{%
(B-A)e^{y}}{(1+e^{y})^{2}}
\end{equation}%
and (\ref{dens1}) becomes 
\begin{equation}
g(y;\theta )=f(\frac{A+Be^{y}}{1+e^{y}};\theta )\cdot \frac{(B-A)e^{y}}{%
(1+e^{y})^{2}}
\end{equation}

We can now find the parameter estimates $\widehat{\theta }$ of $\theta $ via
maximum likelihood. The log-likelihood function $l(\theta )$ is given by 
\begin{equation}
l(\theta )=\sum_{i=1}^{n}\log g(X_{i}^{T};\theta )=\sum_{i=1}^{n}\log
f(X_{i};\theta )\cdot \frac{(B-A)e^{X_{i}^{T}}}{(1+e^{X_{i}^{T}})^{2}}
\end{equation}%
and we can see that it has a rather simple form.

One small problem with the transformation approach is that it makes
visualization of the fit somewhat difficult. Of course we would like to
visualize the fit on the original data. To achieve this we need to
\textquotedblleft back-transform\textquotedblright\ the estimated density.
If $\widehat{g}(y)$ is the density estimate of the transformed data at the
point $y$, then the corresponding estimate of the original data at the point 
$x=H^{-1}(y)$ is given by%
\begin{equation}
\widehat{f}(x)=\widehat{g}(H(x))\cdot \frac{d}{dx}H(x)=\widehat{g}%
(H(x))\cdot \frac{B-A}{(x-A)(B-x)}
\end{equation}%
and we see that this function has singularities at the boundary points $A$
and $B$, even though it still is a proper density which integrates out to $1$%
. Because the fitting is done entirely in the transformed space this does
not affect the parameter estimates.

An added benefit to this approach is that the transformed data becomes more
\textquotedblleft Gaussian-like\textquotedblright\ and as a consequence the
formula for the bandwith $h$ (Eq. \ref{bandwidth}) is known to be close to
optimal. The use of transformations for this purpose was discussed in
Hartert \cite{Hartert}.

\section{Semiparametric Fitting}

The idea of mixing parametric and nonparametric components in a model is
quite common in regression problems but has received fairly little attention
in density estimation. For some work in the statistics literature see Olkin
and Spiegelman \cite{Olkin}, Schuster and Yakowitz \cite{Schuster} and
Faraway \cite{Faraway}.

In HEP numerous papers have used the idea of estimating the background
non-parametrically, though in general without a consideration of the
boundary problem. Cranmer \cite{Cranmer3} presents a general discussion of
KDE and suggests using the boundary kernel method or the reflection method
in case of a boundary. Adelman \cite{Adelman} and Meyer \cite{Meyer} are two
examples of the implementation of the boundary kernel method in a HEP
analysis. Our solution to the boundary problem, though, is more generally
applicable and easier to use.

Say we have a set of observations (or events) $X_{1},..,X_{n}$ with $%
X_{i}\in \lbrack A,B]$. We assume that each event was either generated by a
background distribution with density $f_{B}(x)$, or by a signal distribution
with density $f_{S}(x;\theta )$. Furthermore, we assume that $(1-\alpha
)100\%$ of the events come from the background distribution. Finally, we
also have a sample of pure background events $Y_{1},..,Y_{m}$.

Notice that the background distribution might also depend on some
parameters, but because we will not use the functional form of $f_{B}$ we do
not need to worry about them. Then the density of the mixing distribution is
given by 
\begin{equation}
f(x;\alpha ,\theta )=(1-\alpha )f_{B}(x)+\alpha f_{S}(x;\theta )
\end{equation}

We can use the maximum likelihood method to find the estimates $\widehat{%
\alpha }$ and $\widehat{\theta }$ of $\alpha $ and $\theta $ by maximizing
the log-likelihood $l(\alpha ,\theta )$ given by 
\begin{equation}
l(\alpha ,\theta )=\sum_{i=1}^{n}\log f(X_{i};\alpha ,\theta )
\end{equation}%
In our situation we don't really know the functional form of $f_{B}$, and we
would therefore like to replace $f_{B}$ by the nonparametric kernel density
estimator. To do this we first have to transform the data to get new
observations $X_{1}^{T},..,X_{n}^{T}$ and $Y_{1}^{T},..,Y_{m}^{T}$ as
described above. Next we use the background sample to find the optimal
bandwidth $h_{opt}$ and then we estimate the background density $\widehat{g}%
_{B}$ at the points $X_{1}^{T},..,X_{n}^{T}$ using $h_{opt}$.

The log-likelihood function then becomes 
\begin{equation}
l(\alpha ,\theta )=\sum_{i=1}^{n}\log \left\{ (1-\alpha )\widehat{g}%
_{B}(X_{i}^{T})+\alpha f_{S}(X_{i};\theta )\cdot \frac{(B-A)e^{X_{i}^{T}}}{%
(1+e^{X_{i}^{T}})^{2}}\right\}
\end{equation}

The maximization could now be done using MINUIT \cite{Minuit} or any other
optimization routine.

If we want to do a hypothesis test for the presence of a signal $%
H_{0}:\alpha =0$ vs $H_{a}:\alpha >0$ we can use the likelihood ratio test
with test statistic

\begin{equation}
T=2\left( l(\widehat{\alpha },\widehat{\theta })-l(0,0)\right)
\end{equation}%
Of course, the conditions of Wilks' theorem \cite{Wilks} are not satisfied
here. Nevertheless, quite often $T$ will have an approximate chi-square
distribution with $1$ degree of freedom anyway. Also, because it is a large
sample theorem, even when it does apply one needs to check whether the
asymptotic regime has been reached. This issue of course applies to the
parametric method just as much as to the semiparametric one.

If instead we want to find a confidence interval or an upper bound for the
parameters, we can find error estimates from the usual large sample theory
of maximum likelihood estimators (Casella and Berger \cite{Casella}). These
worked quite well in the simulation studies we conducted. As an alternative
one could find the error estimates via the statistical bootstrap (Efron \cite%
{Efron}). An example of the use of kernel density estimation as well as the
bootstrap method in Astrophysics is discussed in Adriani \cite{Adriani}.

\section{Case Study}

As an illustration, consider the following: The background is itself a 50-50
mixture of a Beta distribution with shape parameters $(1,10)$ and a uniform
distribution. This density covers two extreme cases: on the left boundary
the true density has a large slope and on the right it has slope $0$. The
signal distribution is a Gaussian $\mu =0.35$ and $\sigma =0.02$.

Figure 1 shows one histogram of $1000$ pure background events and another of 
$1000$ data events of which $50$ are signal events. The other two histograms
correspond to these same two samples but with the variable transformed. For
the transformed histograms, their respective semiparametric fits are shown
as continuous curves. Here we fit for both the number of signal events and
the signal location. The estimates are $51.0\pm 9.8$ and $0.354\pm 0.005$.
(The estimates using the the exact parametrization would have been $52.9\pm
9.8$ and $0.353\pm 0.005$). If instead we had used the bootstrap to estimate
errors we would have found $51.0\pm 10.1$ and $0.354\pm 0.005$. Figure 2
shows the results of a sensitivity study. Using the statistical bootstrap
and $5$ different bandwidth selection methods we find that the true optimal
bandwidth is between $0.42$ and $0.52$. Varying $h$ over this range we see
that neither the estimates nor their errors depend on the exact choice of
bandwidth $h$.

\section{Simulation Study}

We will consider six different cases:

\textbf{Case 1 }background is a 80-20 mixture of a Beta(3,20) and a
Beta(3,6). In this case the density goes to 0 as x goes to 0 or 1, so no
transformation is necessary.

\textbf{Case 2 }background is an exponential distribution with rate 1,
truncated at x=1. An example of a density with a moderate slope on one side
(x=0)

\textbf{Case 3 }background is a 50-50 mixture of a Beta(1,10) and a uniform
[0,1]. An example of a density with large slope on one end and slope 0 on
the other.

\textbf{Case 4 }background is a uniform on [0,1]

\textbf{Case 5 }background is an exponential rate 1. An example with a
boundary on just one side. For this we use the transformation $Y=\log (X)$.

\textbf{Case 6 }an example of a multivariate problem. The background is a
bivariate normal with means $(0,0)$, standard deviations $(0.3,0.5)$
correlation $0$ and truncated to $[0,1]^{2}$.

In cases 1-5 a signal is modeled as a Gaussian mean $0.35$ and standard
deviation $0.02$. In case 6 it is a bivariate Gaussian with means $%
(0.35,0.35)$ standard deviations $(0.02,0.02)$ and correlation $0$.

Examples are shown in figures 1 (1000 background events) and figure 2 (950
background events and 50 signal events).%

\begin{center}
  \includegraphics[width=0.90\textwidth]{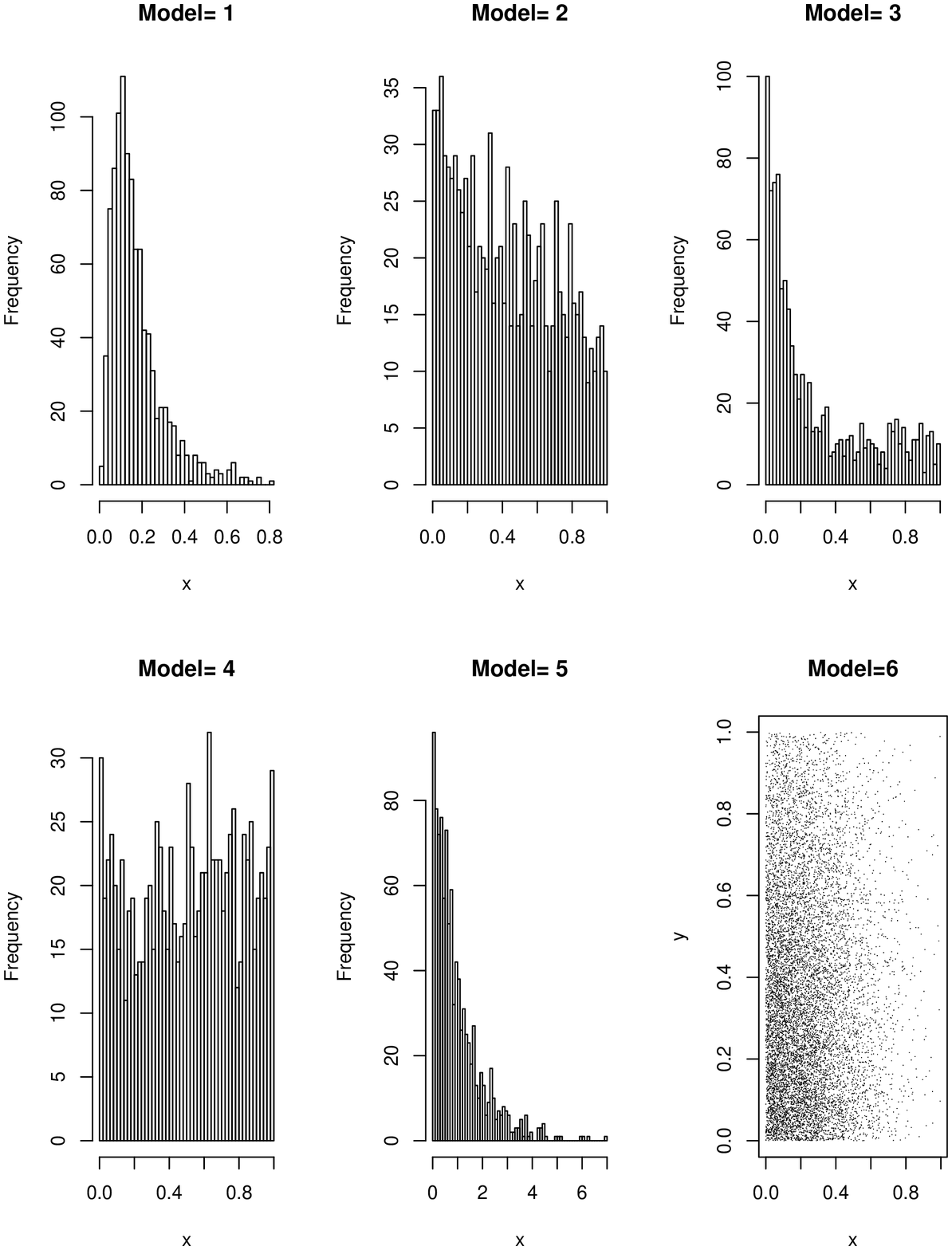}
\end{center}

\begin{center}
  \includegraphics[width=0.90\textwidth]{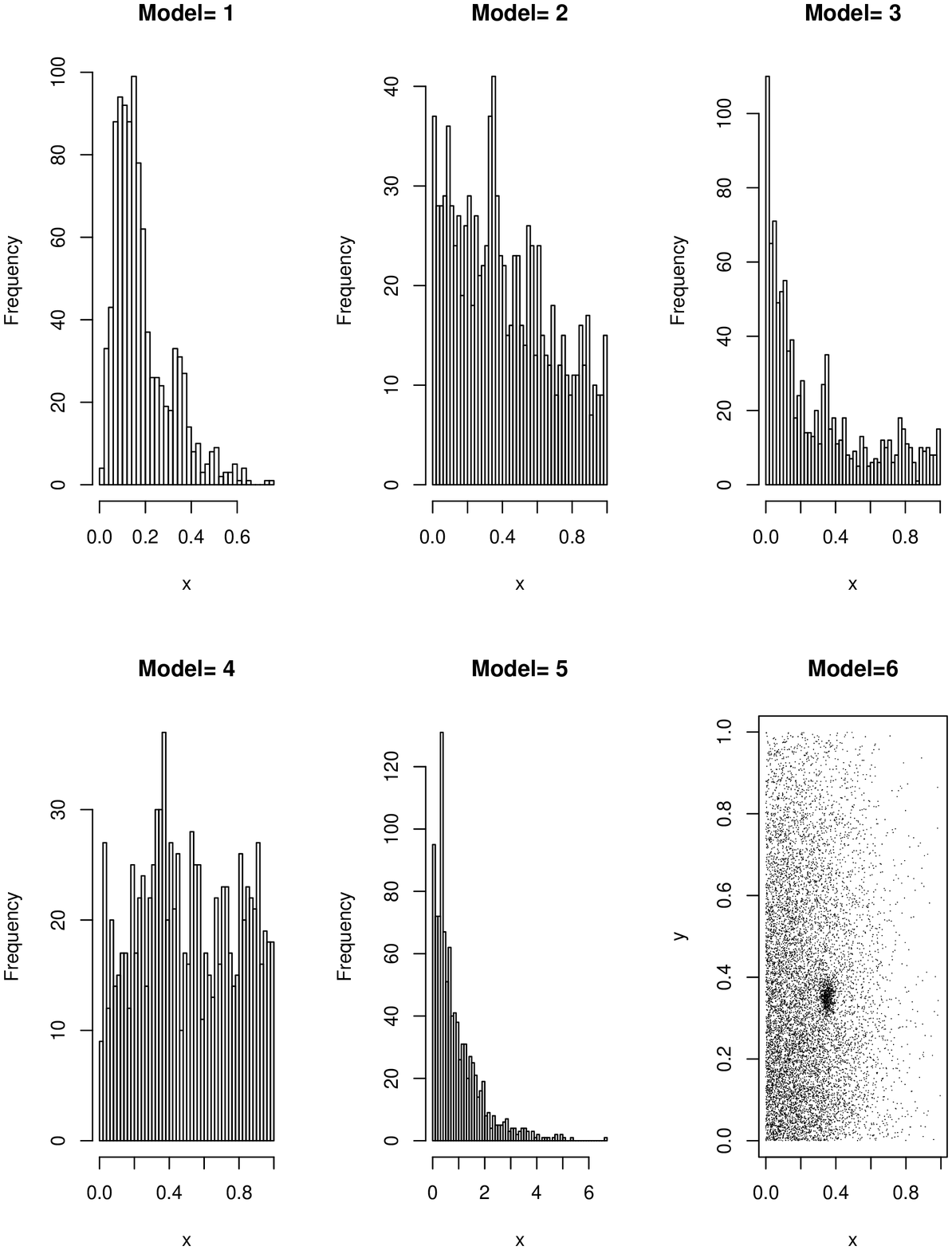}
\end{center}

We begin with the test for the presence of a signal. Here for the method to
work it has to achieve the nominal type I error probability $\alpha $. We
will use $\alpha =1\%$ in our study which means we require that the
estimated number of signal events be positive and that the likelihood ratio
test statistic be larger than $5.4$. We generate $1000\tau $ events each for
the pure background sample and $1000$ events of the "data sample", that is
we assume a ratio of $\tau $ between the sizes for the sideband and the
signal region. We use $\tau =1/2$, $1$, $2$ and $4$. Because we are testing
the null there is no signal present. We find the likelihood ratio statistic
using the three background estimates described above, except \ for model 6
where we include only the exact parametrization and the semiparametric fit.
Here the signal location and width are fixed so there is no issue of the
look-elsewhere effect. This is repeated $25000$ times. The curve for the
exact parametrization is in blue, for the semiparametric method in green and
for the false parametrization in red. We find%

\begin{center}
  \includegraphics[width=0.90\textwidth]{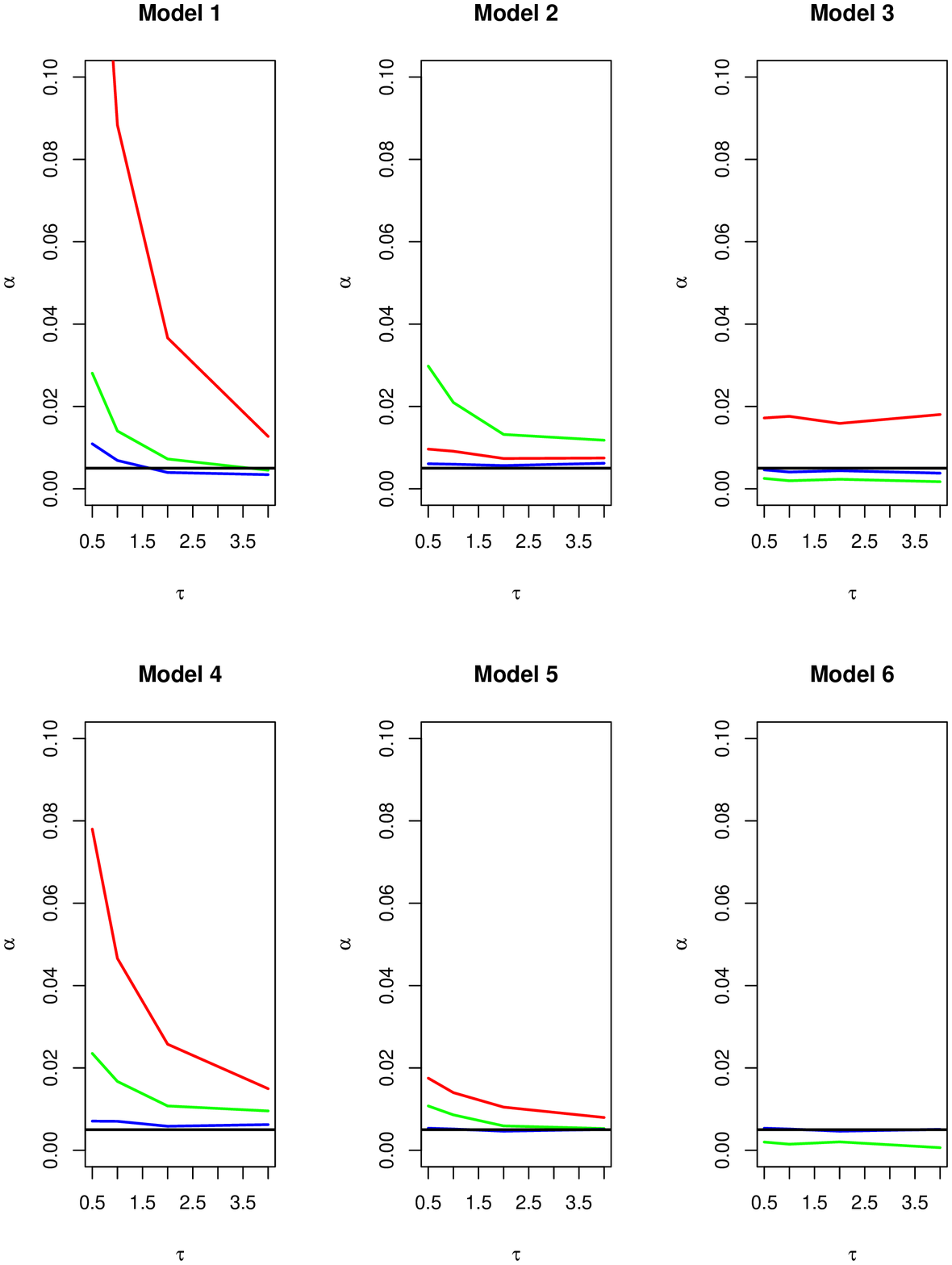}
\end{center}

We see that using the exact parametrization the true type I error
probability is just about the nominal one. Using the semiparametric method
the true type I error probability is sometimes a little higher than the
nominal one but approaches it for larger background samples. Using the
slightly false parametrization it can be as much as 10 times the nomial one.

Next we add a signal drawn from a Gaussian distribution $\mu =0.35$ and $%
\sigma =0.02$. Both data sets have $1000$ events. We vary $\alpha $ from $%
0.5\%$ $\ $to $5\%$, equivalent to $5$ to $50$ signal events. Each
simulation run is repeated $2500$ times.

First, what is the power of the test, that is its ability to detect the
signal?%

\begin{center}
  \includegraphics[width=0.90\textwidth]{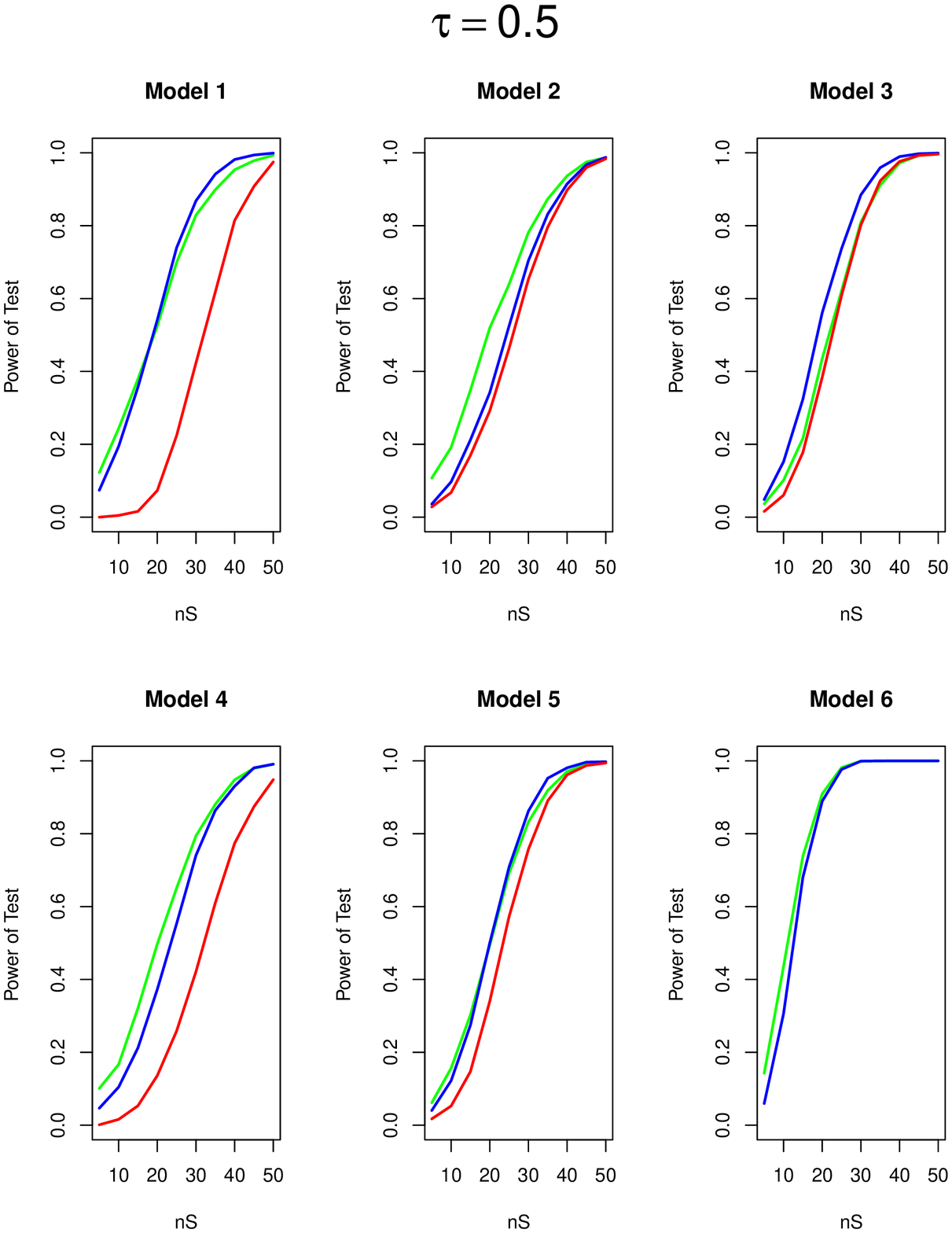}
\end{center}

\begin{center}
  \includegraphics[width=0.90\textwidth]{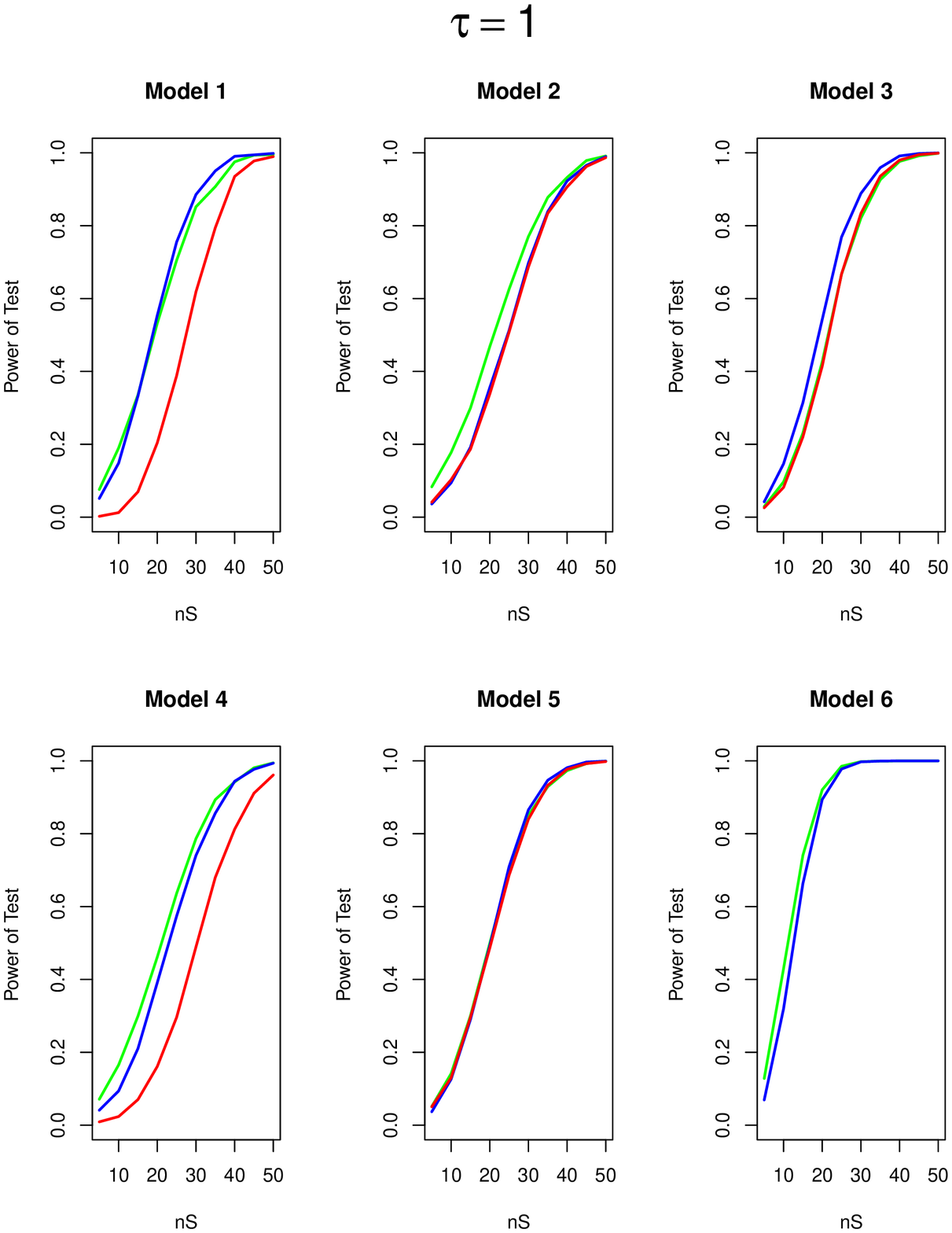}
\end{center}

\begin{center}
  \includegraphics[width=0.90\textwidth]{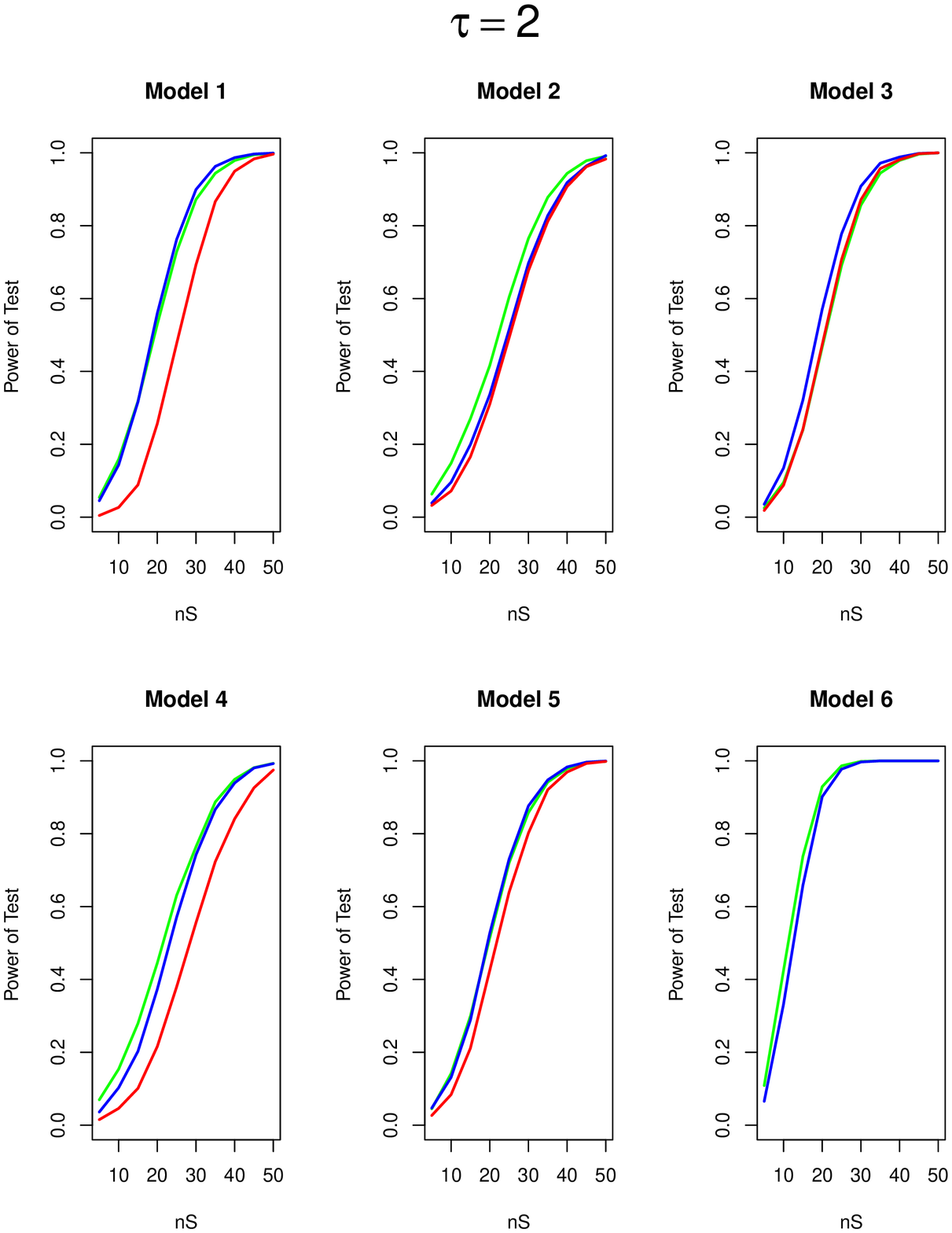}
\end{center}

\begin{center}
  \includegraphics[width=0.90\textwidth]{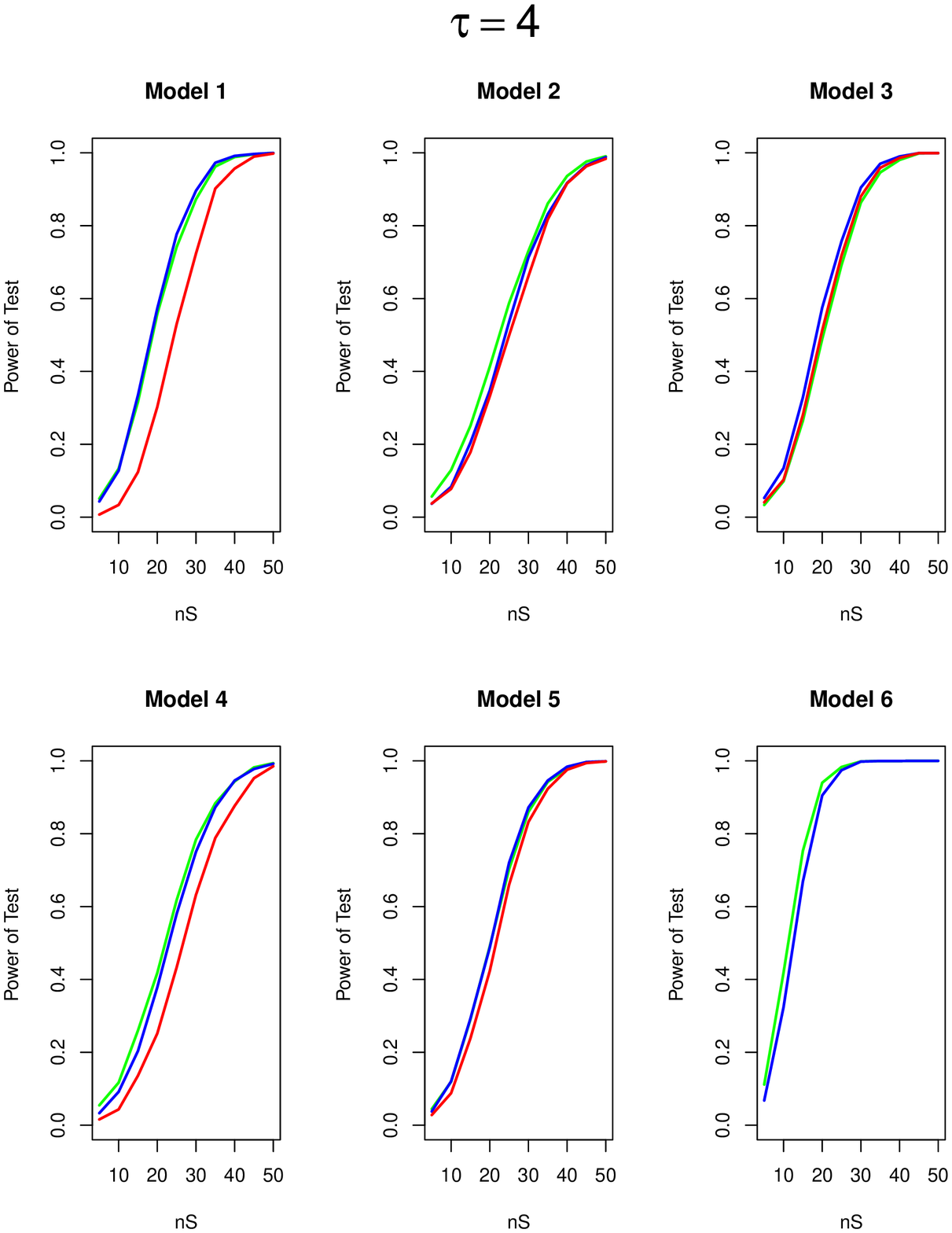}
\end{center}

The power of the test based on the semiparametric fit is almost as good as
the true parametric fit. Note that the power of the "false" parametrization
is missleading because it already had a type I error that was to large.

Are they estimating the signal sizes correctly? Here are the means of the MC
estimates:

\begin{center}
  \includegraphics[width=0.90\textwidth]{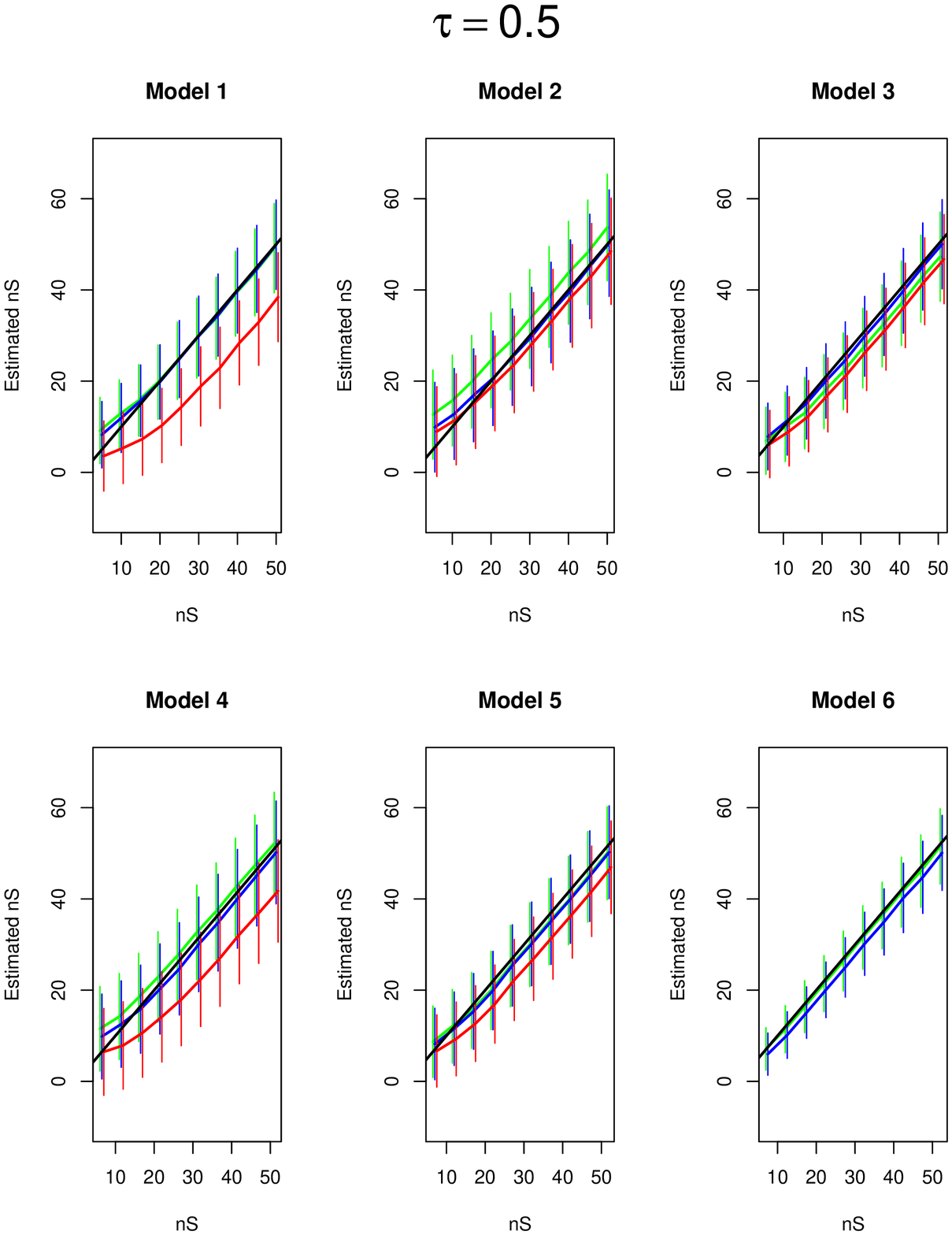}
\end{center}

\begin{center}
  \includegraphics[width=0.90\textwidth]{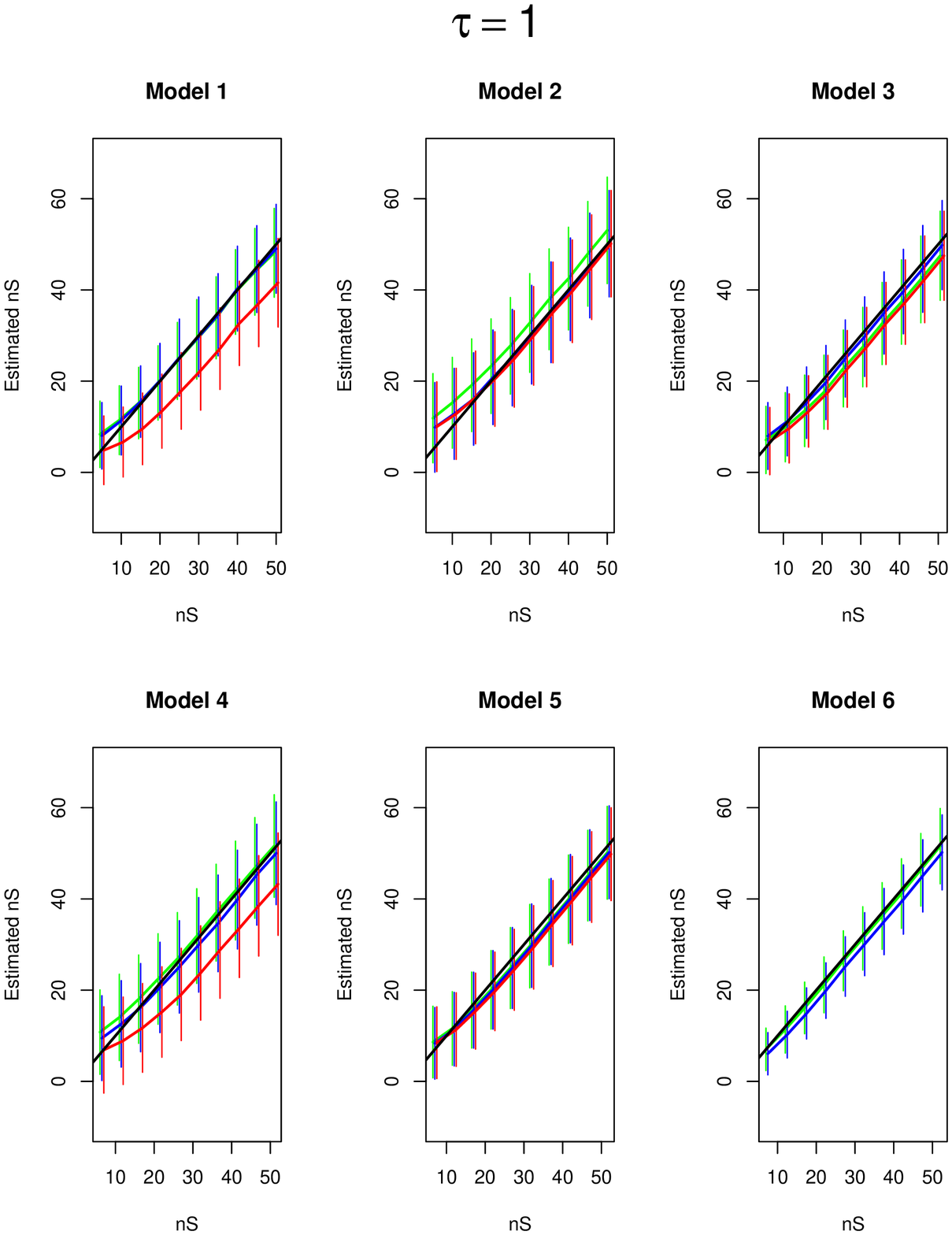}
\end{center}

\begin{center}
  \includegraphics[width=0.90\textwidth]{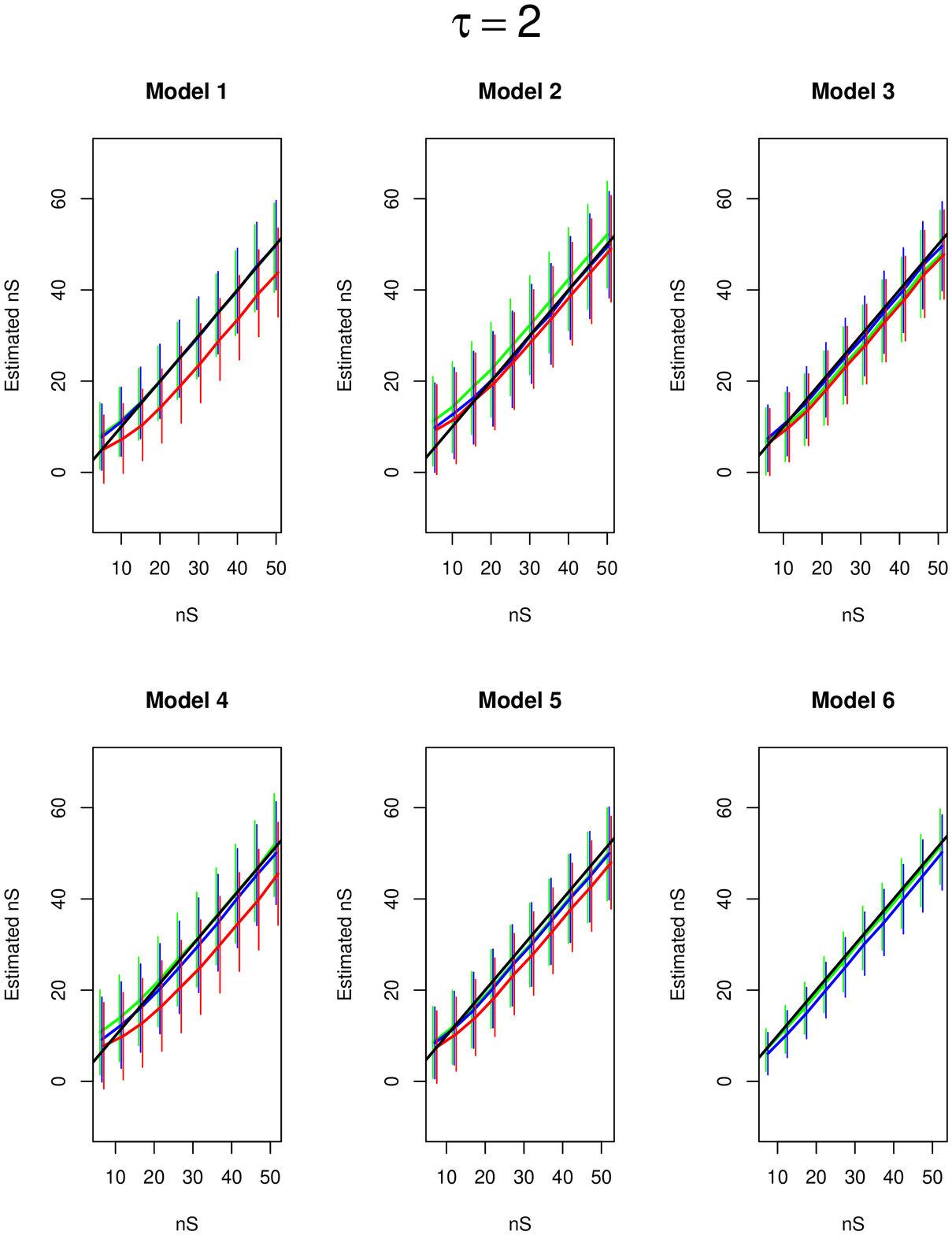}
\end{center}

\begin{center}
  \includegraphics[width=0.90\textwidth]{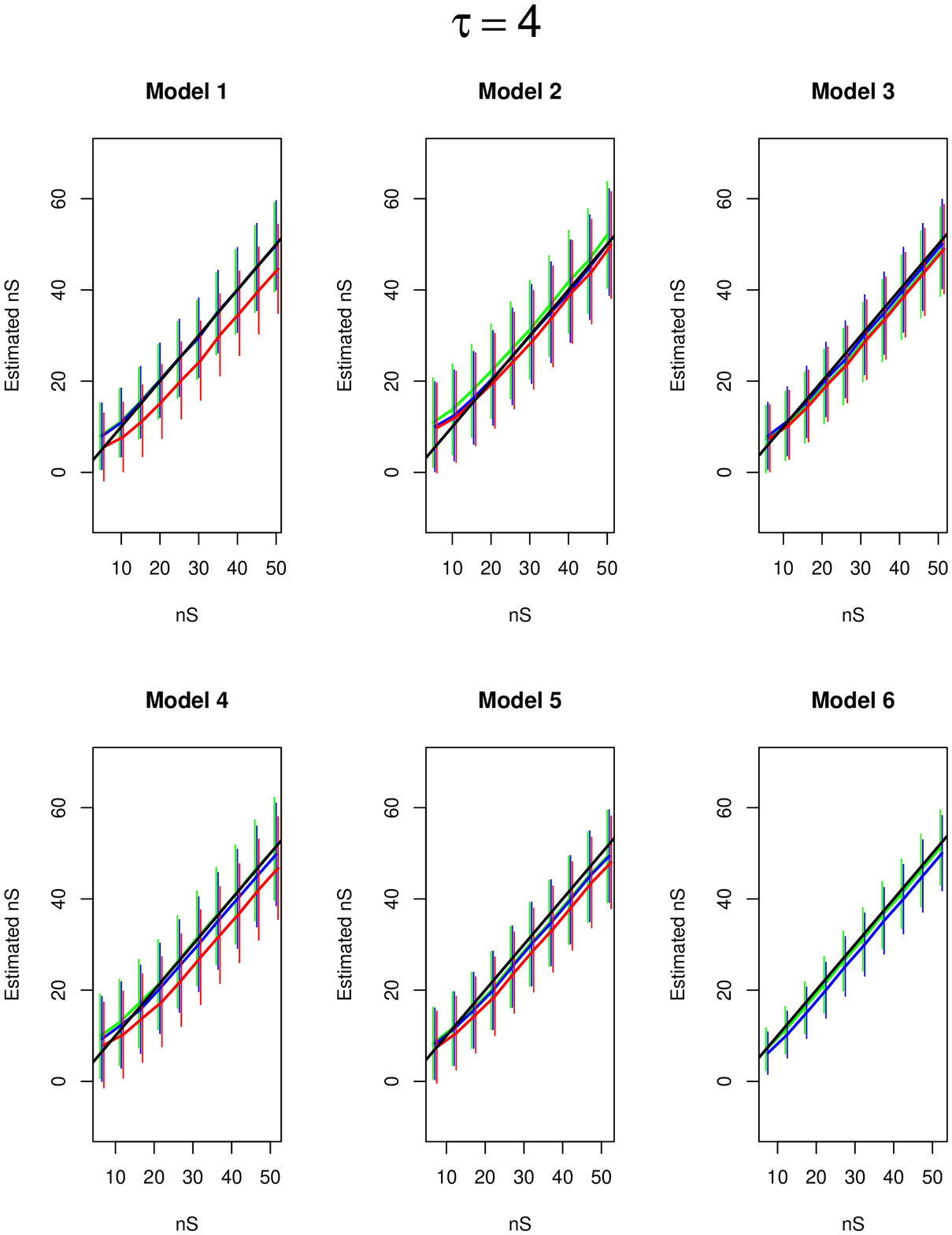}
\end{center}

Are the error estimates correct? For this we will calculate the 1 standard
deviation confidence intervals and check whether the percentage of intervals
containing the true number of signal events is the required $68\%$:%

\begin{center}
  \includegraphics[width=0.90\textwidth]{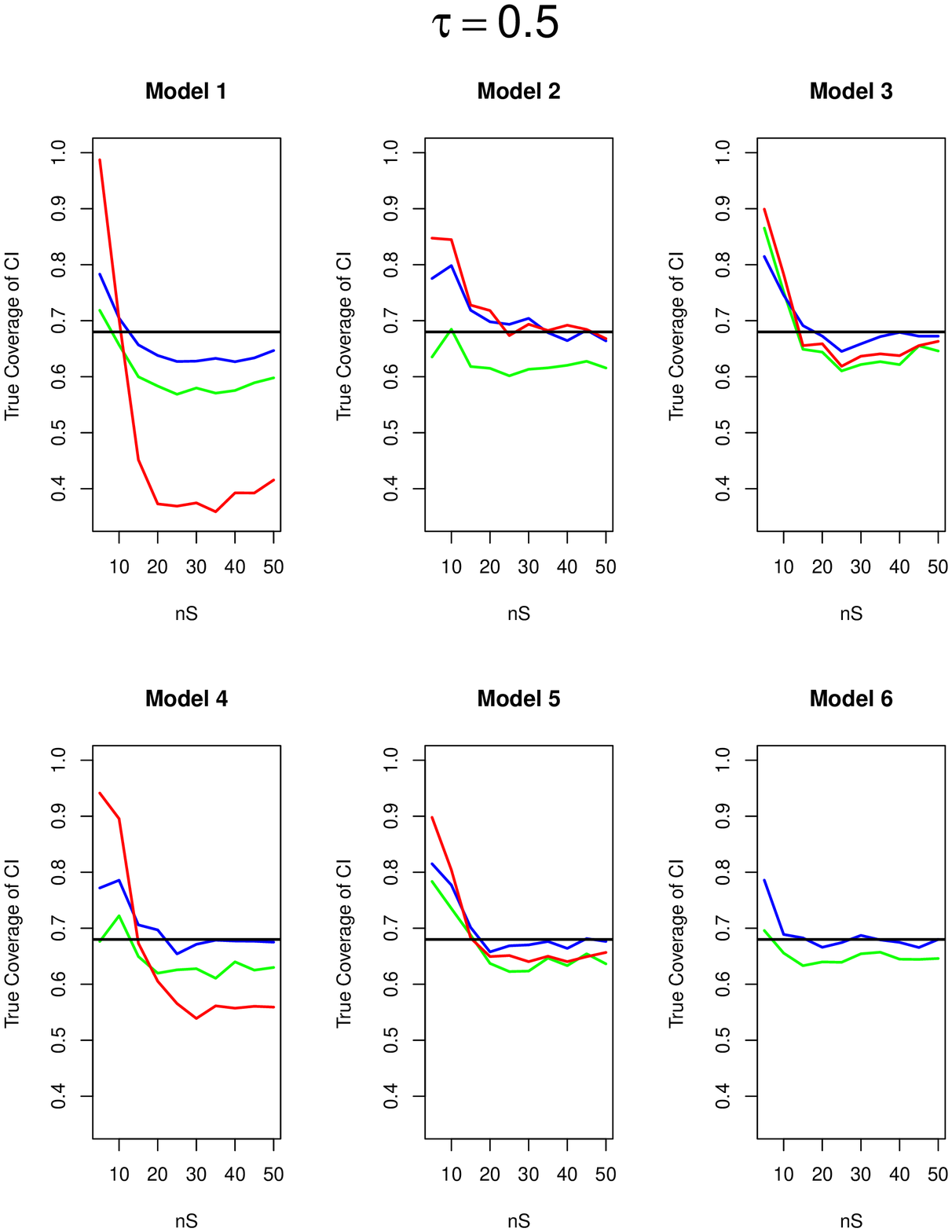}
\end{center}

\begin{center}
  \includegraphics[width=0.90\textwidth]{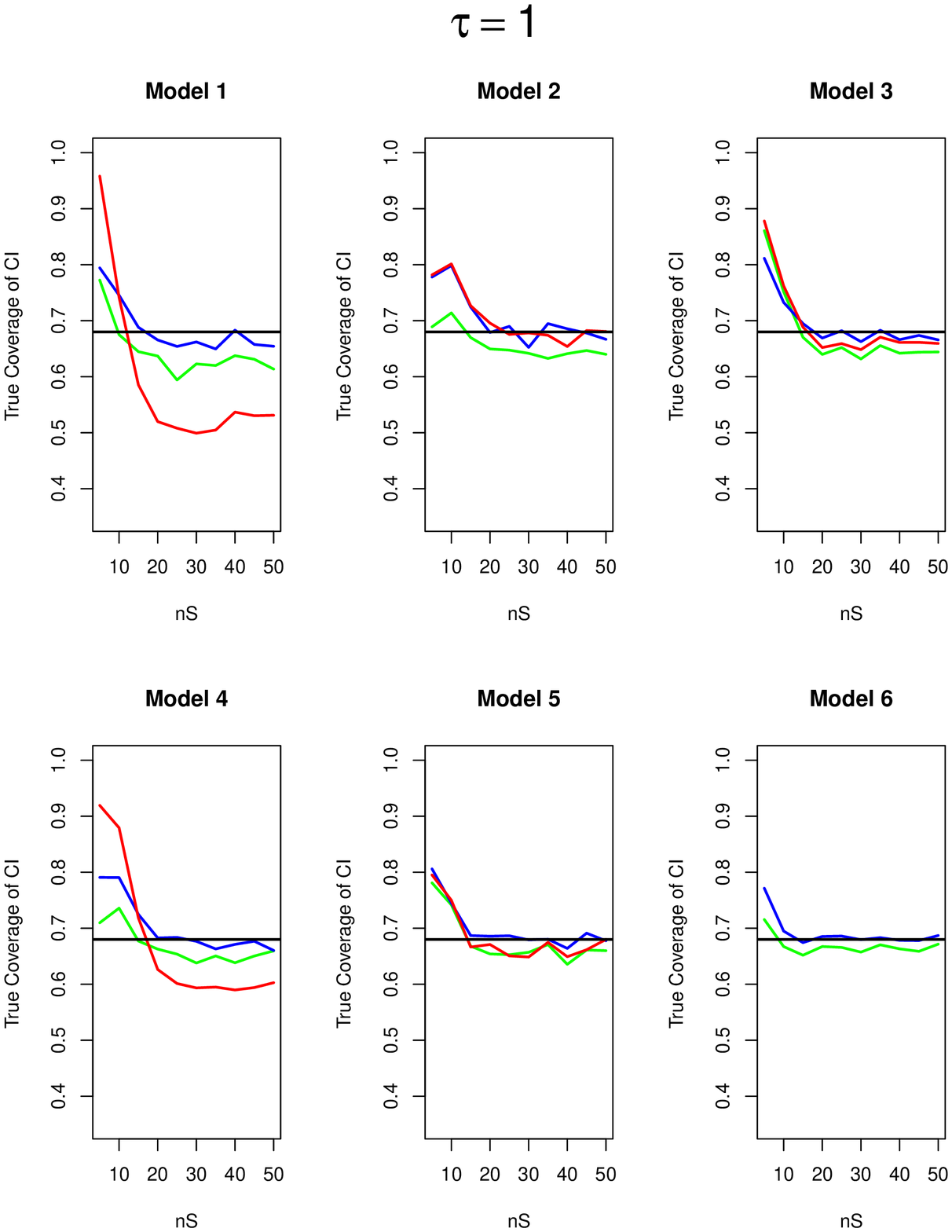}
\end{center}

\begin{center}
  \includegraphics[width=0.90\textwidth]{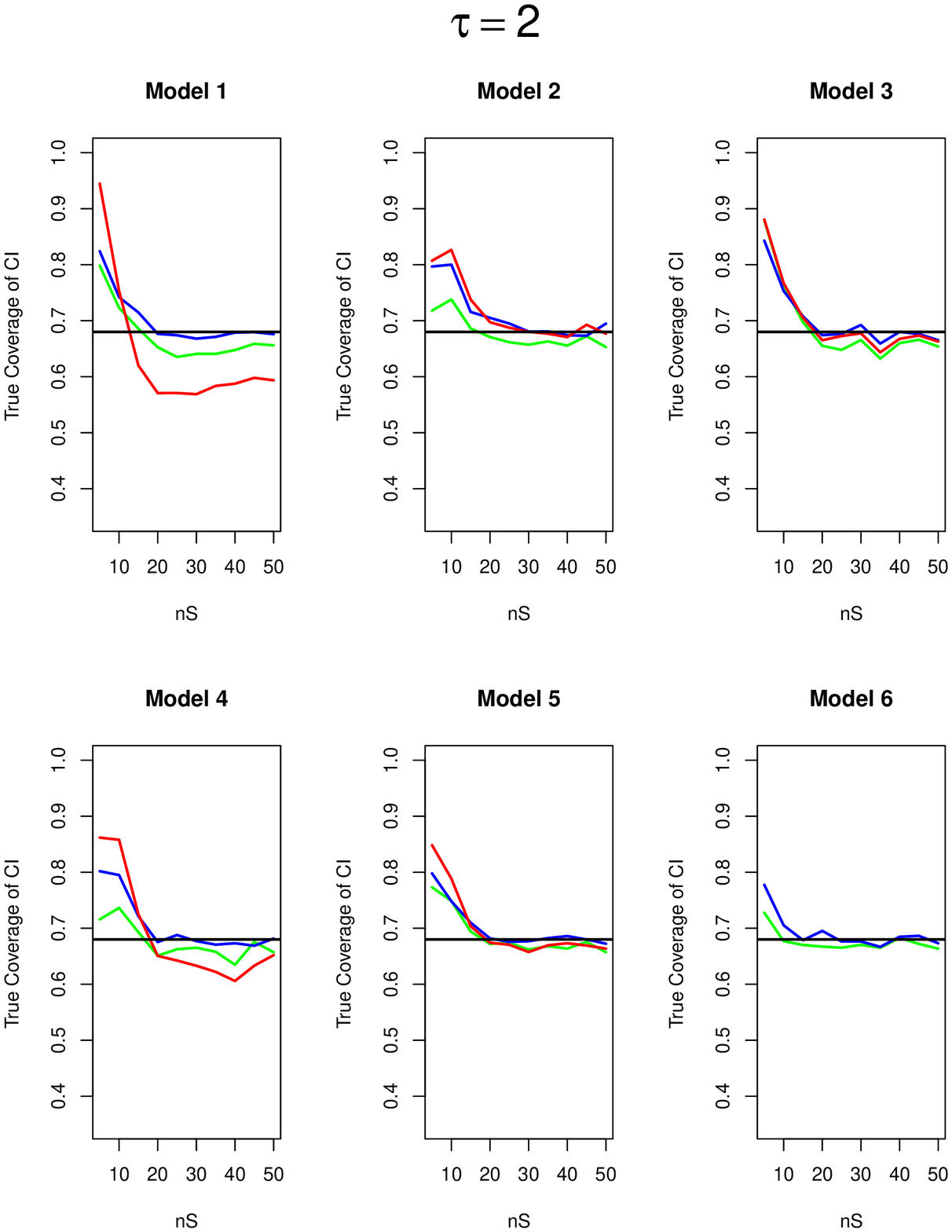}
\end{center}

\begin{center}
  \includegraphics[width=0.90\textwidth]{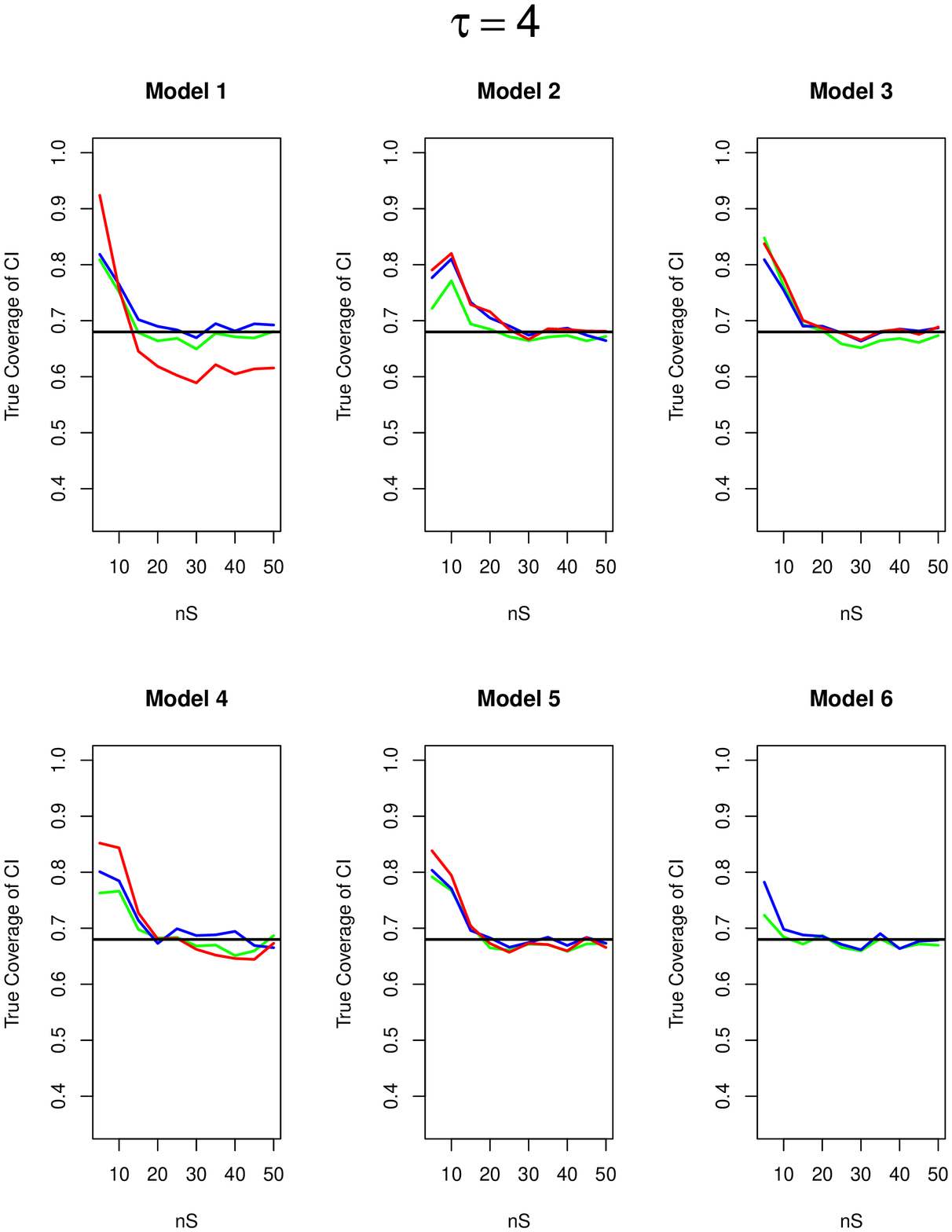}
\end{center}

Again the semiparametric method has errors just about correct, whereas the
false parametrization can yield errors quite wrong.

Generally the semiparametric method does slightly worse than the exact
parametrization (which unfortunately in real life is unknown) but better
than a slightly false but "reasonably" looking one.

\section{Computational Issues}

The work described in this paper was done using the statistics package R,
available at http://cran.r-project.org/, which includes the routine \emph{%
density} for finding the optimal bandwidth $h$ (with a choice of 5 different
methods). This routine was also used to calculate the density estimates. R
uses function calls to C++ routines which are freely available. The HEP
analysis platform RooFit, available at http://roofit.sourceforge.net/,
includes the KEYS program written by Kyle Cranmer. In Perl, an
implementation can be found in the Statistics-KernelEstimation module. In
Java, the Weka package, available at http://www.cs.waikato.ac.nz/ml/weka/,
provides weka.estimators.KernelEstimator, among others. In Gnuplot, kernel
density estimation is implemented by the smooth kdensity option; the data
file can contain weight and bandwidth for each point, or the bandwidth can
be set automatically. C++ code for calculating the optimal bandwidth is
available at http://www.umiacs.umd.edu/labs/cvl/pirl/vikas and is described
in Raykar \cite{Raykar}. Recently developed algorithms for the so-called
Fast Gauss Transform have made it possible to calculate the NPD estimates
even for very large data sets. One example is discussed in Yang et al. \cite%
{Yang1}. A sample C routine which calculates limits for Model 2 using the
semiparametric and the correct parametric background is available from the
authors at http://charma.uprm.edu/$\sim$rolke/publications.htm .

\section{Conclusions}

We describe a method that uses nonparametric density estimation to estimate
the background density and then finds estimates and errors for the
parameters of the signal such as the number of signal events via
maximum-likelihood. This is made feasible by the availability of a pure
background sample. A simulation study shows that this method is quite
competitive, almost as good as using the (in real life unknown) true
parametrization and superior to using an almost correct parametrization.
Moreover, because the nonparametric density estimate depends smoothly on the
bandwidth, it is easy to do a sensitivity study and gain insight into the
systematic uncertainty caused by different background shapes.

\section{Acknowledgement}
This material is based upon work funded by the U.S. Department of Energy, Office of Science, Office of High Energy Physics under Award Number DE-FG-02-97ER41045 and by the University of Puerto Rico.

\end{document}